\documentclass{ptapap}
\usepackage{amsmath}
\usepackage{amssymb}

\usepackage{color}
\newcommand{\annot}[1]{{\textbf{\textcolor{red}{#1}}}}
\usepackage{soul}

\author{Dominika {\L}. Kr\'ol}[CFT,UJ]
\author{Agnieszka Janiuk}[CFT]
\affil[CFT] {Center for Theoretical Physics PAS, Al. Lotnikow 32/46, 02-668 Warsaw, Poland }
\affil[UJ]{Astronomical Observatory, Jagiellonian University, ul. Orla 171, 30-244 Krak\'ow, Poland}
\title{Evolution of black hole mass and spin in collapsars}

\begin{document}
	
\maketitle

\begin{abstract}
We investigate the collapsar scenario for the long gamma ray bursts.
The energetics 
in the $\gamma$-ray band are consistent with
the binding energy of a progenitor star. The events duration times, lightcurve profiles, variability, and connection with supernovae, are still subject of many studies. In our scenario, the
evolved progenitor star is collapsing onto a spinning
black hole formed from the compact core. The accretion via rotationally supported torus powers
the ejection of relativistic jets. The rotational energy of the black hole is
presumably transported to the remote jet and mediated by magnetic fields.
The rotation of pre-supernova star is a key property of the model. In our study, we investigate different  accretion scenarios and  various distributions of initial angular momentum in the envelope, that affect the spinning up the black hole and its mass increment.
\end{abstract}
	
	
	
\section{Tested pre-supernova models}
	
We used pre-supernova models from \citet{Woosley02}, \citet{Heger00}, and \citet{Heger05} with $M_{\text{ZAMS}}\sim25 M_{\odot}$. First of them (NR) similarly to the one used in 
\citet{Janiuk08a} and \citet{Janiuk08b} did not take into account rotation\annot{,} neither 
magnetic field 
during star evolution. Its initial metallicity was $10^{-4}$ solar. There is no mass-loss in this 
case. In the second star (R) rotation was introduced, hence this model is affected by the biggest 
mass-loss leading to pre-supernova mass of $5.62 M_{\odot}$. In the third model (RMF) both rotation 
and magnetic field is incorporated in evolution. It leads to pre-supernova mass of $12.55 
M_{\odot}$. For the second and third model initial metallicity was equal to solar.
	
\section{Models of accretion}

Critical specific angular momentum $l_{\rm crit}$  is given by the equation:\newline
$l_{\rm crit}=\frac{2GM}{c}\sqrt{2-A+2\sqrt{1-A}}$, where $M$ and $A$ denote mass and spin of black 
hole. Torus can be sustained when there is a matter with $l_{\rm spec}>l_{\rm crit}$.The original 
distribution of angular momentum in the stellar envelope was given by $l_{\rm spec}=\text{x}l_{\rm 
crit}f(r,\theta),$ where $f(r,\theta)$ depends on the model. We examined variety of x values.
Every step of black hole mass (M) and angular momentum (J) evolution in our model is given by 
following set of equations: 
\begin{equation}
M^{\text{k}}=M^{\text{k-1}}+\Delta m
\end{equation}
\begin{equation}
J^{\text{k}}=J^{\text{k-1}}+\Delta j
\end{equation}
\begin{equation}
\Delta 
m^{\text{k}}=2\pi\int^{r_{k}+\delta r}_{r_{k}}\int^{\pi}_{0}\rho r^{2}\text{sin}\theta d\theta 
dr\quad
\end{equation}
\begin{equation}
\Delta j^{\text{k}}=2\pi\int^{r_{k}+\delta r}_{r_{k}}\int^{\pi}_{0}min[l_{\rm spec},l_{\rm
crit}]\rho r^{2}\text{sin}\theta d\theta dr.
\end{equation}

\noindent Thereafter spin parameter $A$ is expressed by the formula: $A=\frac{cJ}{G(M)^{2}}$.
This evolution formula ensure fulfillment of the condition $A<1$, but do not specify any mechanism 
of the angular momentum loss. We investigate two accretion scenarios (scenario 1 and 3 accordingly 
from \citet{Janiuk08a}). Accretion proceeds through: both torus and the envelope in the first 
scenario, only torus in the second. Calculations were stopped when there was no matter with  $l_{\rm 
spec} > l_{\rm crit}$. We present examination of four models, with combinations of accretion 
scenarios and $f(r,\theta)$ formula:
$f(r,\theta)=1-|\text{cos}\theta|$ \textbf{(A models)} and 
$f(r,\theta)=\text{sin}^{2}\theta\sqrt{\frac{r}{r_{in}}}$ \textbf{(D models)}.
	
\subsection{Models with angular momentum distribution given by function A}

\begin{figure}[t]
\includegraphics[width=\textwidth]{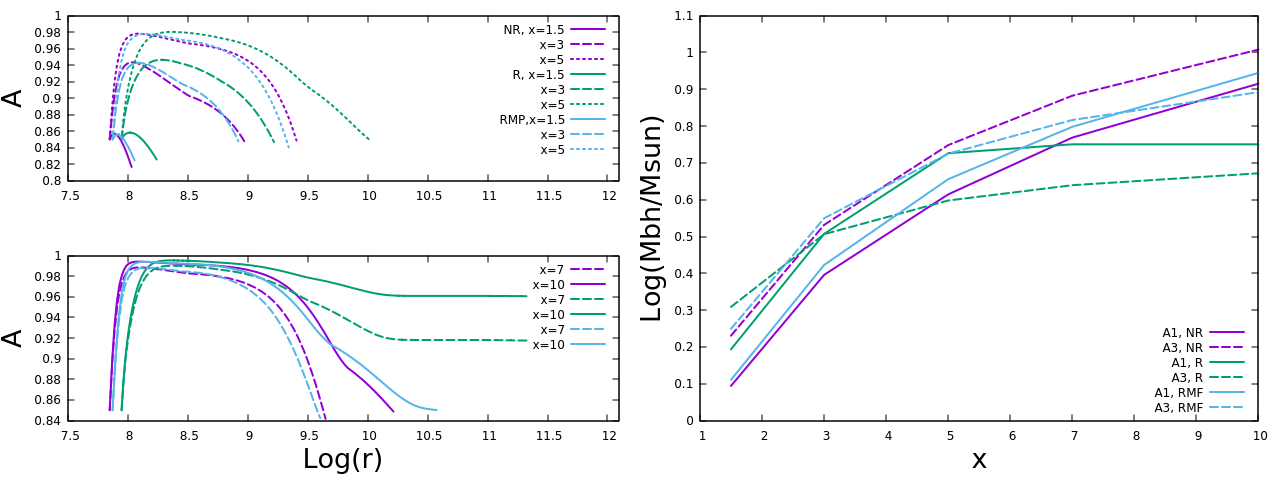}
\caption{Left panel: evolution of the spin parameter for different pre-spernova 
models and values of $x$ parameter. Right panel: final mass of the 
black hole as a function of $x$ parameter.}\label{A1}
\end{figure}

In the left panel of Fig. \ref{A1} we present evolution of the spin parameter for the A1 model. The 
evolution depends heavily on the pre-supernova star type. Accretion in RMF and NR models ends at the 
same radii. For small $x$ their final $A$ is similar, but with increase of $x$ the 
difference between 
final spins grows, RMF leads to higher values. Pre-supernova model R allows the support of 
torus for 
higher r. Starting from $\log{r}\sim10.5$ accretion almost doesn't change $A$, 
due to the low density in outer most parts of the pre-supernova.  For A3 scenario (i.e., only torus 
accretion), all examined $x$ values lead to rapid growth of $A$ to value $\sim1$ and maintaining 
this value until the end of the accretion.
In the right panel of Fig. \ref{A1} we present final mass of the black hole as a function of $x$. 
Higher $x$, which corresponds to higher specific angular momentum of the matter, leads to more 
massive black holes, because accretion in our model can proceed longer.
	
\subsection{Models with angular momentum distribution given by function D}
	
In case of model D1, we examined different range of $x$ values. Due to dependence of $l_{\rm spec}$ 
on $\sqrt{r}$, the $x$ values from A1 model lead to very high initial $l_{\rm spec}$. Therefore, now 
we examined values in the range $x=[0.05, 1.0]$. 
	
\begin{figure}
\includegraphics[width=\textwidth]{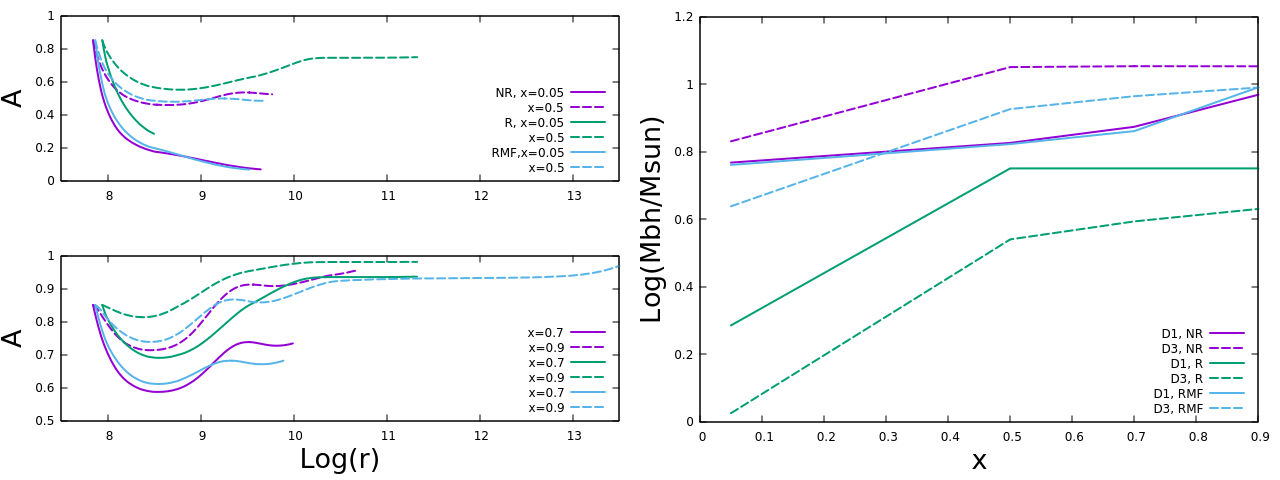}
\caption{Left panel: evolution of the spin parameter for different pre-spernova models and values of $x$ parameter. Right panel: final mass of the black hole as a function of $x$ parameter.}
\label{D1}
\end{figure}

Dependence of the spin of BH on pre-supernova model is similar to the A1 model. For small $x$ 
values, the NR and RMF models are terminated after roughly the same part of the star's envelope has 
accreted, and they give similar final $A$ values. Situation changes for faster rotation, with 
$x=0.9$. For this $x$ value, both NR and RMF models sustain the torus until the outermost part of 
the star starts accrete, and the difference in the radius of termination occurs due to 
different 
sizes of the stars.
In case of pre-supernova model with rotation, the whole star accretes for $x$ larger than 0.5. 
Evolution of the Kerr parameter is presented in the left panel of Fig. \ref{D1}. Final mass of the 
black hole as a function of $x$ is shown in the right panel of Fig. \ref{D1} 
	
\section{Summary}
	
Final black hole masses are in the range of $\log(\frac{M_{\rm BH}}{M_{\odot}}) = [0.6 - 1.05]$,  
depending on the accretion scenario and pre-supernova model.
In case of spin evolution through consecutive accretion layers, models A1 and D1 give different 
shape of the $A$ evolution:
in case of A1 scenario for every pre-supernova model we observe growth at first and then depending 
on exact model and $x$ value, the spin may drop or saturate, at the outer parts of 
the star.
On the other hand, every D1 model gives drop of $A$ at first and then growth for high enough value 
of $x$.  We acknowledge support by the grant no. DEC-2016/23/B/ST9/03114 from the Polish National Science Center.

\bibliographystyle{ptapap}
\bibliography{krol.bib}

\end{document}